\documentclass[12pt,reqno]{article}

\usepackage{amsmath}
\usepackage{amsthm}
\usepackage{amssymb}
\input{xy}
\xyoption{all}

\newtheorem{lemma}{Lemma}
\newtheorem{proposition}{Proposition}

\DeclareMathOperator{\sign}{\text{sign}}
\def\S{{\cal S}}
\def\c{\gamma} 

\def\a{\alpha}
\def\b{\beta} 
\def\c{\gamma} 
\def\d{\delta} 
\def\e{\epsilon} 

\def\R{{\cal R}}

\newcommand{\eq}{\begin{eqnarray*}}
\newcommand{\qe}{\end{eqnarray*}}
\newcommand{\eqn}{\begin{eqnarray}}
\newcommand{\qen}{\end{eqnarray}}

\newcommand{\Real}{\mathbb{R}}
\newcommand{\Complex}{\mathbb{C}}

\newcommand{\Integer}{\mathbb{Z}}
\newcommand{\mat}{\begin{pmatrix}}
\newcommand{\tam}{\end{pmatrix}}

\newcommand{\mc}{\mathcal}
\newcommand{\mf}{\mathfrak}
\newcommand{\p}[2]{\{{#1},{#2}\}} 
\def\op{\oplus}

\newcommand{\gl}{{\mf{gl}}}
\newcommand{\sll}{{\mf{sl}}}
\newcommand{\psl}{{\mf{psl}}}

\def\g{\mathfrak{g}}
\def\h{\mathfrak{h}}

\def\ti{\otimes}
\def\sfrac12{{\scriptstyle \frac12}}
\def\tfrac32{{\scriptstyle \frac32}}

\title{\bf Tensor products of $\psl(2|2)$ representations}  
\author{\\[5mm] Gerhard G\"otz$^1$, Thomas Quella$^2$, 
               Volker Schomerus$^{1,3}$ \\[5mm] 
$^1$ Service de Physique Th\'eorique, CEA Saclay,\\ 
F-91191 Gif-sur-Yvette, France\\[5mm] 
$^2$ Department of Mathematics, King's College London, \\ 
Strand, London WC2R 2LS, UK\\[5mm]
$^3$ DESY Theory Group, Notkestrasse 85,\\
D-22603 Hamburg, Germany}
\date{} 
\begin{document}
\begin{titlepage}      \maketitle       \thispagestyle{empty}

\vskip1cm
\begin{abstract} 
The aim of this work is to study finite dimensional representations of 
the Lie superalgebra $\psl(2|2)$ and their tensor products. In particular, 
we shall decompose all tensor products involving typical (long) and atypical 
(short) representations as well as their so-called projective covers. While 
tensor products of long multiplets and projective covers close among 
themselves, we shall find an infinite family of new indecomposables in 
the tensor products of two short multiplets. Our note concludes with a 
few remarks on possible applications to the construction of $AdS_3$ 
backgrounds in string theory.  
\end{abstract} 

\vspace*{-17.9cm}
\noindent
{\tt {SPhT-T05/080}} \hfill {\tt DESY 05-092} \\ 
{\tt {KCL-MTH-05-05}} \hfill {\tt hep-th/0506072}\\
\bigskip\vfill
\noindent

{\small e-mail: }{\small\tt
Gerhard.Goetz@cea.fr \& quella@mth.kcl.ac.uk \& 
vschomer@mail.desy.de }
\end{titlepage} 

\baselineskip=19pt 
\setcounter{equation}{0} 
\section{Introduction}
\def\tr{{\rm tr}}
\def\PSL{{PSL(2$|$2)}}
\def\SL2R{{SL(2,$\mathbb{R}$)}}
\def\SL2C{{SL(2,$\mathbb{C}$)}}
\def\SU2{{SU(2)}}
\def\sl2{{sl(2)}}

This short note is devoted to the representation theory of the 
Lie superalgebra $\psl(2|2)$. The latter describes symmetries 
of several important physical systems, ranging from strings 
moving in an $AdS_3$ background \cite{Rahmfeld:1998zn,Berkovits:1999im} to 
the quantum Hall effect (see e.g.\ \cite{Zirnbauer:1999ua}). 
The questions we address here, however, are purely mathematical 
and we shall only comment very briefly on our motivation from 
physics, leaving more concrete applications to a future 
publication. 
\smallskip 

The so-called A-series of simple Lie superalgebras
\cite{Kac:1977em} consists of 
$\sll(n|m)$ with $n\neq m$ and $\psl(n|n)$.\footnote{The Lie 
superalgebras $\psl(n|n)$ are obtained from $\sll(n|n)$ when we 
remove the 1-dimensional center.} Finite dimensional 
representations of these Lie superalgebras with diagonal Cartan
elements have 
been constructed and investigated extensively (see e.g.\ 
\cite{Kac1977:MR0444725,Scheunert:1977wj,Kac1978:MR519631} for some early papers). 
Most of this work focused on irreducible representations, 
disregarding that even the very simplest Lie superalgebras 
from the A-series possess plenty of non-trivial indecomposables 
(see e.g.\ \cite{Scheunert:1977wj,Marcu:1979se}). 
In fact, the Lie superalgebras $\sll(n|m)$ with $n,m>1$ admit 
so many of them  
that they cannot even be classified \cite{Germoni1998:MR1659915}. 
Irreducible representations, on the other hand, are rather easy 
to list (see e.g.\ \cite{Frappat:1996pb} and references therein). 
These fall into two different classes known as {\em typical} 
(long) and {\em atypical} (short) representations \cite{Kac1978:MR519631}. 
More general indecomposables may be regarded as composites of 
the latter.% 
\smallskip%

Investigations of tensor products for the A-series cannot 
avoid dealing with
indecomposable representations. In fact, it is well known 
that the product of two irreducibles is often not fully 
reducible \cite{Scheunert:1977wj,Marcu:1979sg} and what 
is even worse: 
indecomposables of Lie superalgebras do not form an ideal 
in the fusion ring. Consequently, it would e.g.\ not be 
possible to determine how many times a given irreducible 
representation appears in a higher tensor product if 
we only knew the number of irreducibles in the two-fold 
products of irreducible representations, simply because 
the fusion paths leading to irreducible representations 
may pass through indecomposables. In other words, the 
study of tensor products for representations of Lie 
superalgebras cannot be consistently truncated to 
irreducible representations. General results show, however, 
that there is a preferable class of so-called projective 
representations which gives rise to an ideal in the fusion 
ring. This contains the typical representations along with 
certain maximal indecomposable composites of atypicals. 
The latter are known as {\em projective covers} of 
atypical representations.%
\smallskip% 

The indecomposable representations that emerge in products 
of irreducibles add a lot of interesting novel structure 
to the study of tensor products, but they certainly also 
bring a lot of extra difficulties. Even though tensor 
products of Lie superalgebras from the A-series have been 
investigated in the literature (see e.g.\ 
\cite{Marcu:1979sg,Hurni:1985vk} 
for some early work), most of the existing work excludes 
degenerate cases in which indecomposables appear as one of 
the factors or in the decomposition of the product. More 
extensive results on degenerate products seem to be restricted 
to a few simple examples, including $\sll(2|1)$ \cite{Marcu:%
1979sg} and $\gl(1|1)$ (see e.g.\ \cite{Rozansky:1992td}). 
And even in these cases, a complete treatment was only 
found recently \cite{Gotz:2005jz}.%  
\smallskip%

In this note we shall study all finite dimensional tensor 
products for $\psl(2|2)$ between typical and atypical 
representations as well as their projective covers. We 
shall see explicitly how the products involving at least 
one projective representation can be decomposed into a sum 
of typicals and projective covers (propositions 1, 2 and 3 below). 
On the other hand, a new family of indecomposables  arises 
in tensor products of atypical representations (see 
proposition 4). The next section contains an introduction
to the three most important types of $\psl(2|2)$ representations. 
Section 3 is then devoted to our new results on tensor 
products of these representations. A complete list of 
our findings is provided in subsection 3.1 along with 
a few additional comments. Our claims are supported by 
two central ideas which are explained in subsection 
3.2 before we demonstrate how they work together in 
our computation of tensor products. Since the full 
calculations are quite cumbersome and not very 
illuminating, we shall illustrate the key steps 
in two representative examples rather than attempting 
to present a general proof. In the concluding section, 
we will discuss at least briefly our motivations from 
physics and some potential applications.

\section{Representations of $\psl(2|2)$} 
\setcounter{equation}{0}

In this first section we shall mainly review known results about 
some finite dimensional representations of $\g=\psl(2|2)$. 
We shall provide a complete list of irreducible representations,
both typical and atypical, explain how they are constructed from 
the so-called Kac modules and we shall describe the projective 
covers of all atypical representations. 

\subsection{The Lie superalgebra $\psl(2|2)$}

The Lie superalgebra $\g=\psl(2|2)$ possesses six bosonic generators $K^{ab} 
= - K^{ba}$ with $a,b= 1,\dots,4$. In addition, there are eight fermionic 
generators that we denote by $S^a_\a$ with the new index $\a = 1,2$ and 
where $a$ assumes the same values as for the bosonic generators. The
relations of $\g$  are given by 
\begin{equation}
\begin{split} 
[K^{ab},K^{cd}] &\ =\ i \bigl[ \d^{ac}K^{bd}-\d^{bc}K^{ad}- 
                        \d^{ad}K^{bc}+\d^{bd}K^{ac}\bigr] \\[2mm]
[K^{ab},S^c_\c] &\ =\ i \bigl[ \d^{ac} S^b_\c - \d^{bc} S^a_\c\bigr] \\[2mm] 
[S^a_\a,S^b_\b] &\ =\ \frac12 \e_{\a\b}\, \e^{abcd} K^{cd} \ \ . 
\end{split}
\end{equation} 
Here, $\e_{\a\b}$ and $\e^{abcd}$ denote the usual $\e$-symbols with 
two and four indices, respectively, and a summation over repeated indices is
implied. 
\smallskip 

Note that the even subalgebra $\g^{(0)} = \psl(2|2)^{(0)}$ is isomorphic to
$\sll(2) \oplus \sll(2)$. The odd part $\g^{(1)}$ of $\psl(2|2)$  is 
spanned by the eight fermionic generators. We split the latter into two 
sets of four generators  
$$ \g^{(1)}_+ =  {\text{span}}\{S_2^a\}\ \ \ , \ \ \ 
   \g^{(1)}_- =  {\text{span}}\{S_1^a\}\ \ . 
$$ 
As indicated by the subscript $\pm$, we shall think of the fermionic 
generators $S_1^a$ as annihilation operators and of $S_2^a$ as creation 
operators. 
\smallskip 

Let us furthermore recall that the group \SL2C acts on $\g$ through 
outer automorphisms. For an element $u = ( {u_{\a}}^{\b}) \in $ 
 \SL2C the latter read  
\begin{equation}
\label{eq:Auto}
\c_u (K^{ab}) \ = \ K^{ab} \ \ , \ \ 
\c_u (S^a_\a) \ = \ {u_{\a}}^{\b}\,  S^a_\b \ \ , 
\end{equation}
Consistency with the defining relations of our Lie superalgebra 
is straightforward to check. It only uses the fact that
$\text{det}(u)=1$. 
 
\subsection{Finite dimensional irreducible representations} 

The irreducible finite dimensional representations of $\g$ are 
labeled by pairs $j_1,j_2$ with $j_i = 0,1/2,1,\dots$. All these 
representations are highest weight representations and they are 
uniquely characterized by the highest weights $(j_1,j_2)$ of the 
corresponding even subalgebra $\g^{(0)} \cong \sll(2) \oplus 
\sll(2)$. We shall see in a moment how such representations 
may be constructed explicitly.   
\smallskip 

For reasons that we shall understand below, the irreducible  
representations of $\g$ fall into two classes. So-called {\em 
typical representations} appear for $j_1\neq j_2$. We shall 
denote them as $[j_1,j_2]$.  Their dimension is given by 
\begin{equation} \label{dtyp} 
 {\rm dim}\ [j_1,j_2] \ = \ d_{[j_1,j_2]} \ = 
    \ 16 \, (2j_1+1)(2j_2+1) \ \ .
\end{equation} 
Representations with labels $j=j_1=j_2$ are {\em atypical}, 
since their dimension is smaller than a naive application 
of formula \eqref{dtyp} would suggest. For these irreducible 
representations we shall employ the symbol $[j]$ and one 
finds that  
$$ {\rm dim}\ [j]\ = \ d_{[j]}  \ = \ 
   16 j (j+1) + 2 \ \ . $$ 
The formula holds for $j \neq 0$. The representation $[0]$ 
is the trivial one-dimensional representation. Let us also 
point out that the 14-dimensional atypical representation 
$[1/2]$ is the adjoint representation of $\psl(2|2)$ . 
\smallskip 

The irreducible representations $[i,j]$ and $[j]$ of the Lie 
superalgebra $\g$  can be restricted to the even subalgebra 
$\g^{(0)}$. With respect to this restricted action they 
decompose according to 
\begin{eqnarray}\label{evendec1} 
[j]\bigr|_{\g^{(0)}} & \cong & (j+\sfrac12,j-\sfrac12) \oplus 2(j,j) 
             \oplus (j-\sfrac12,j+\sfrac12) \qquad (\text{for }j>0)\ \ ,
             \\[2mm] \label{evendec2} 
[i,j]\bigr|_{\g^{(0)}} & \cong & (i,j) \otimes \bigl[2(0,0) \oplus 
   2(\sfrac12,\sfrac12) \oplus (0,1)\oplus (1,0)\bigr] \ \ .  
\end{eqnarray} 
Here and in the following, the pairs $(i,j)$ denote irreducible 
representations of the even subalgebra. Note that these 
decomposition formulas are consistent with our expressions
for the dimension of irreducible representations. 
\smallskip 

The irreducible representations of $\g$  possess one property that 
will become very important later on: They admit an implementation  
of the outer automorphisms $\c_u$, see \eqref{eq:Auto}. To make
this more precise, let 
us introduce the symbol $\pi$ for the representation that sends  
elements $X \in \psl(2|2)$ to linear maps on the representation 
space ${\cal H}_\pi$. If we compose a representation $\pi$ with 
an automorphism $\c$ of the Lie superalgebra, we obtain a new 
representation $\pi \circ \c$ on the same graded vector space. In 
general, this new representation differs from $\pi$. If $\pi$ is 
one of our finite dimensional irreducible representations $[i,j]$
or $[j]$ and $\c$ one of the automorphisms in \eqref{eq:Auto},
however, then the new representation is equivalent to 
the original one, i.e.\ for every $u \in$ \SL2C there exists an 
invertible linear map 
$U_{\pi}: {\cal H}_{\pi} \longrightarrow 
{\cal H}_{\pi}$ such that 
\begin{equation}
  \label{eq:Imp}
  \pi \circ \c_u (X) \ = \ 
    U_{\pi} \ \pi (X)\  U_{\pi}^{-1} \ \ \ \text{ for all } \ \ \ X \in \psl(2|2)\ \ . 
\end{equation} 
Let us stress that the map $u \rightarrow U_{\pi}$ defines a 
representation of the subgroup \SL2C of outer automorphisms on 
the representation space ${\cal H}_{\pi}$.

\subsection{Kac modules and irreducible representations} 

It is useful for us to discuss briefly how the irreducible 
representations we have listed above can be constructed.
The idea is rather standard: We begin with an irreducible 
highest weight representation $(j_1,j_2)$ of the even 
subalgebra $\g^{(0)}$. We declare that the corresponding 
representation space $V_{(j_1,j_2)}$ is annihilated by $S_1^a$ 
and then generate a so-called Kac module $[j_1,j_2]$ by
application of the raising operators $S^a_2$, 
$$  [j_1,j_2] \ := \ 
    {\rm Ind}_{\g^{(0)} \oplus\, \g^{(1)}_-}^{\g} V_{(j_1,j_2)} \ = \ 
    \mc{U}(\g) \otimes_{\g^{(0)} \oplus\, \g^{(1)}_-} V_{(j_1,j_2)} \ \ .    
$$ 
Here, we have extended the $\g^{(0)}$ module $V_{(j_1,j_2)}$ 
to a representation of $\g^{(0)} \oplus \, \g_-$ by setting $S^a_1 
V_{(j_1,j_2)} = 0$. Note that we can apply at most four 
fermionic generators to the states in $V_{(j_1,j_2)}$. Therefore, 
the dimension of this Kac module is given 
by 
$${\rm dim}\ [j_1,j_2] \ = \ 16 (2j_1+1) (2j_2+1) 
\ \ . $$ 
The Kac module $[j_1,j_2]$ is irreducible whenever $j_1 \neq 
j_2$. In these generic cases it agrees with the typical 
representation. For $j_1 = j_2= j$, however, the associated 
Kac module turns out to be reducible but not fully reducible, 
i.e.\ it cannot be written as a direct sum of irreducible 
representations. If $j \geq 1$ the structure of the 
Kac module can be encoded in the following chain 
\begin{equation}  
 [j,j]: \ [j] \ \rightarrow \  [j+\sfrac12] 
   \oplus [j-\sfrac12]\ \rightarrow [j] \ \ ,  
\end{equation}  
or, equivalently, in a planar diagram in which one direction 
refers to the spin $j$ of the atypical constituents, 
\begin{equation} \label{Kj}
 [j,j]: \  \xymatrix{ & [j+\sfrac12]\ar[dr] &\\
     [j]\ar[dr] \ar[ur] && [j]\ \ . \\
             & [j-\sfrac12] \ar[ur]&}
\end{equation} 
Since pictures of this type will appear frequently throughout 
this text, let us pause here for a moment and explain carefully 
how to decode their information. We read the diagram \eqref{Kj} 
from right to left. The rightmost entry in our chain contains 
the so-called {\em socle} of the indecomposable representation, 
i.e.\ the largest fully reducible invariant submodule we can find. 
In the case of our Kac module, the socle happens to be irreducible 
and it is given by the atypical representation $[j]$. If we 
divide the Kac module by the submodule $[j]$, we obtain a new 
indecomposable representation of our Lie superalgebra. 
Its diagram is obtained from the one above by removing the 
last entry and arrow. The socle of this quotient is a 
direct sum of the two atypical representations $[j\pm 1/2]$. 
It is rather obvious how to iterate this procedure until the 
entire indecomposable representation is split up into floors 
with only direct sums of irreducible representations appearing 
on each floor. 
\smallskip 

There are two special cases for which the decomposition of the 
Kac module does not follow the generic pattern. These are the 
cases $j = 0$ and $j=1/2$, 
\begin{eqnarray} 
 [0,0]: &  & [0] \ \rightarrow \  [\sfrac12] 
    \rightarrow [0] \ \ ,\label{K0} \\[2mm]
 [\sfrac12,\sfrac12]: &  & [\sfrac12] \ \rightarrow \  
 [1] \ \rightarrow \ [0]\oplus [0] \ \rightarrow 
 [\sfrac12] \ \ .  \label{K12}  
 \end{eqnarray} 
Let us note that our formula for the dimension of atypical 
representations follows  directly from the decomposition of 
the corresponding Kac modules.      
  
\subsection{Projective covers of atypical representations} 
\def\P{{\cal P}}

As we have seen in the last subsection, atypical representations
can be extended into larger indecomposables. Kac modules are only 
one example of such composites and we shall indeed see several 
others as we proceed. Among them, however, there is one special
class, the so-called projective covers ${\cal P}_\g(j)$. By definition, 
these are the largest indecomposables whose socle consists of a 
single atypical representation $[j]$. General results imply that 
such a maximal indecomposable extension of $[j]$ exists and is 
unique. In case of $j \geq 3/2$, the  structure of  
$\P_\g(j)$ is encoded in the following diagram 
\begin{eqnarray} \label{Pj} 
{\cal P}_\g(j):  & & [j] \longrightarrow 
             2 [j+\sfrac12] \oplus 2 [j-\sfrac12]
                \longrightarrow  
              [j+1]\oplus 4[j] \oplus 
               [j-1] \longrightarrow \\[2mm] 
           & & \hspace*{4cm} \longrightarrow 
                2 [j+\sfrac12] \oplus 2 [j-\sfrac12]
               \longrightarrow [j] \ \ . \nonumber 
\end{eqnarray}
Note that $\P_\g(j)$ contains an entire Kac module as proper 
submodule. In this sense, the Kac modules are extendable. We 
also observe one rather generic feature of projective covers: 
they are built up from different Kac modules in a way that 
resembles the pattern in which Kac modules are constructed 
out of irreducibles (see eq.\ (\ref{Kj})). One may see this 
even more clearly if $\P_\g(j)$ is displayed as a 2-dimensional 
diagram in which the additional direction keeps track of 
the spin $j$ of the atypical constituents $[j]$, 
\begin{equation} \P_\g(j): \ 
  \xymatrix{& &  [j+1] \ar[dr]& &\\
 & 2[j+1/2]\ar[dr] \ar[ur] & & 2[j+1/2]\ar[dr]&\\
     [j]\ar[dr] \ar[ur] && 4[j]\ar[ur]\ar[dr] & &[j]\ . \\
             & 2[j-1/2] \ar[ur]\ar[dr]&&2[j-1/2]\ar[ur] &\\
& & [j-1] \ar[ur] &}
\end{equation}
We will continue to switch between such planar pictures and
diagrams of the form \eqref{Pj}. The remaining cases $j=0,1/2,1$ 
have to be listed separately. When $j=1$ the picture is very 
similar only that we have to insert $2[0]$ in place of 
$[j-1]$, 
\begin{equation} \label{P1} 
{\cal P}_\g(1):  \   [1] \longrightarrow 
               2 [{\scriptstyle \frac{3}{2}}] 
                  \oplus 2 [\sfrac12]
               \longrightarrow  [2]\oplus 4[1] \oplus 
               2 [0] \longrightarrow 
                2 [{\scriptstyle \frac{3}{2}}] 
                  \oplus 2 [\sfrac12]
               \longrightarrow [1] \ \ . 
\end{equation}      
The projective cover of the atypical representation $[1/2]$ is 
obtained from the generic case by the formal substitution $2[j-1/2] 
\rightarrow 3[0]$, 
\begin{equation} \label{P12} 
{\cal P}_\g(\sfrac12): \   [\sfrac12] \longrightarrow 
           2 [1] \oplus 3 [0] \longrightarrow
              [{\scriptstyle \frac{3}{2}}] 
                  \oplus 4 [\sfrac12] 
               \longrightarrow   2 [1] \oplus 3 [0]
               \longrightarrow [\sfrac12] \ \ . 
\end{equation} 
Finally, the projective cover ${\cal P}_\g(0)$ of the trivial representation 
is given by, 
\begin{equation} \label{P0} 
{\cal P}_\g({0}):  \  [0] \longrightarrow 3 [\sfrac12]
           \longrightarrow 2 [1] \oplus 6 [0] 
               \longrightarrow [\sfrac12] 
             \longrightarrow [0] \ \ . 
\end{equation} 
The reader is invited to convert the last three formulas into 
planar pictures. This concludes our list of the projective 
covers. The representations
${\cal P}_\g(j)$ will arise as important building blocks in the 
decomposition of tensor products in the next section. Together, 
typical representations and the projective covers of atypicals
form the subset of so-called projective representations. What 
makes this class particularly interesting is its behavior under 
tensor products. In fact, it is rather well-known that projective 
representations of a Lie superalgebra form an ideal in the fusion
ring. We will see this very explicitly in the concrete 
decomposition formulas for tensor products below.

\section{Tensor products of $\psl(2|2)$ representations} 
\setcounter{equation}{0}
\def\P{{\cal P}}

This section contains the central results of this work, i.e.\ 
explicit formulas for the decomposition of tensor products 
between all the finite dimensional irreducible representations
and projectives we have 
introduced in the previous section. We begin by stating 
our main results. Then we shall explain the two key ideas 
that are needed in the proof. Finally, we analyze 
two very representative examples.

\subsection{Tensor products of irreducible representations} 

Before we start listing our results, we would like to 
introduce one object that will enable us to determine the 
contributions from typical representations in most of the 
tensor products below. We shall denote by $\pi_\g$ a map 
which associates a direct sum of typical representations 
to any finite dimensional representation of the bosonic 
subalgebra. On the irreducible representations of $\g^{(0)}$ 
it gives   
\begin{equation}
\pi_\g(i,j) \ : = \ \begin{cases}  [i,j] \ \ 
                      &  \text{ for } \  i \neq j \\[2mm] 
                       0 & \text{ for }  \ i = j \ \ . 
                          \end{cases}    
\end{equation} 
This prescription is extended to any direct sum of irreducibles 
by linearity. $\pi_\g$ enters e.g.\ in the following formula for
the tensor product of typical and atypical representations. 
\bigskip

\begin{proposition}[Tensor product of typical and atypical
representations] The tensor product of an atypical representation 
$[n]$ with a typical representation $[i,j]$ is given by 
$$ [n] \otimes [i,j] \ = \ \pi_\g\bigl([n]\bigr|_{\g^{(0)}} \otimes 
   (i,j)\bigr) \ \ \ \ \text{ for } \ \ \ i+j \ \not\in \ \Integer\ . 
$$
When $i+j \in \Integer$, on the other hand, the tensor product can also 
contain projective covers, 
\begin{equation} 
[n] \otimes [i,j] \ = \ \pi_\g\bigl([n]\bigr|_{\g^{(0)}} \otimes 
   (i,j)\bigr) \oplus \ 
 \bigoplus_{l = p}^{q}\ {\cal P}_\g(l) \ \    
\end{equation} 
where $p = {\rm max}(|n-i|,|n-j|)$ and $q= {\rm min}(n+i,n+j)$. The 
decomposition of the irreducible representation $[n]$ into 
representations of the even subalgebra $\g^{(0)}$ was spelled out 
in eq.\ \eqref{evendec1}.
\end{proposition}
\bigskip 

It is interesting to observe that the tensor product of the 
representation $[n]$ with the typical representation $[0,2n]$ 
contains a single projective cover ${\cal P}_\g(n)$. Proposition
2 may be employed to determine tensor products between typical 
and any indecomposable representation. The prescription requires 
introducing a new map $\S_\g$ which sends an indecomposable 
representation $\R$ to a sum of atypicals, 
\begin{equation}\label{S}  
 {\S_\g}(\R) \ = \ \bigoplus_j \, [\R:[j]] \, \cdot \, [j] \ \ . 
\end{equation}  
Here, the symbol $[\R:[j]]$ denotes the total number of atypical 
representations $[j]$ in the decomposition series of $\R$. In the 
case where $\R$ is one of the projective covers, for example, these
numbers can be read off from eqs.\ \eqref{Pj}-\eqref{P0}. With this 
notation, the tensor product between a typical representation 
$[i,j]$ and the indecomposable $\R$ reads, 
\begin{equation}  \label{typInd} 
\R \otimes [i,j] \ \cong \ \S_\g(\R) \otimes [i,j]  \ \ . 
\end{equation}    
As we have anticipated, we may now employ proposition 1 to 
decompose the tensor product on the right hand side. Thereby, 
we can determine e.g.\ the fusion of typical representations 
and projective covers.   
\bigskip 

\begin{proposition}[Tensor product of typical representations]  
The tensor product of two atypical representations $[i_1,j_1]$ and $[i_2,j_2]$
is given by 
$$ 
[i_1,j_1] \otimes [i_2,j_2]  \ = \ \pi_\g \bigl([i_1,j_1]\bigr|_{\g^{(0)}}
              \otimes (i_2,j_2) \bigr) \ \ \ \ \ \text{ for } \ \ \
      i_1+i_2+j_1+j_2 \ \not\in \ \Integer \ . $$ 
If $i_1+i_2+j_1+j_2 \in \Integer$, on the other hand, projective covers may 
appear in the decomposition,  
\begin{eqnarray} 
[i_1,j_1] \otimes [i_2,j_2]  & = & \pi_\g \bigl([i_1,j_1]_{\g^{(0)}}
              \otimes (i_2,j_2) \bigr) \ \oplus \ 
    2\cdot\!\bigoplus_{m=2p}^{2q} \P_\g({\scriptstyle{\frac{m}{2}}}) \\[2mm]
    & &  \hspace*{-4cm} \oplus \begin{cases}
     \hspace*{4cm}\delta_{p,q+1}\,\P_\g(q+\sfrac12) &,\ p>q\\[2mm]
     \ \bigl(1-\delta^{i_1+i_2}_{j_1+j_2}\bigr)  
    \cdot \P_\g(q+\sfrac12) \ 
    \oplus \ \bigl(1-\delta^{|i_1-i_2|}_{|j_1-j_2|}\bigr) \cdot \P_\g(p-\sfrac12)    
    \ \ominus \ \delta_{p,0} \cdot \P_\g(0) &,\ p\leq q \ \ .
    \nonumber \end{cases}
\end{eqnarray}
The decomposition of typical representations $[i,j]$ into
irreducibles of ${\g^{(0)}}$ appears in eq.\ \eqref{evendec2}. 
We have also introduced the parameters $p$ and $q$ by 
$p = {\rm max}(|i_1-i_2|,|j_1-j_2|)$ and $q = 
{\rm min}(i_1+i_2,j_1+j_2)$. Note 
that the last term in the first line subtracts one copy of 
the projective cover $\P_\g(0)$ whenever $p$ vanishes.
\end{proposition}

At this point we are able to decompose all tensor products which 
involve at least one typical factor. Our next task is to analyze 
the products in which at least one factor is a projective cover 
$\P_\g(j)$. 
\bigskip 

\begin{proposition}[Tensor product of atypical 
representations and projective covers]  The tensor product of 
an atypical representation $[n]$  with a projective cover 
$\P_\g(j)$ with $j\neq 0$ and $n \neq 0$ is given 
by 
\begin{eqnarray} 
[n] \otimes \P_\g(j)  & = & \pi_\g \bigl( [n]\bigr|_{\g^{(0)}} \otimes  
       H_{(j)}\bigr) \ \oplus  \ 2\cdot\!\!\!\bigoplus_{m=2|n-j|}^{2(n+j)} 
      P_\g({\scriptstyle{\frac{m}{2}}}) \ \ominus \ \delta_{n,j} 
  \cdot \P_\g(0) \\[2mm] 
\text{where} & & H_{(j)} \ = \ 2 (j,j) \oplus (j+\sfrac12,j+\sfrac12)
\oplus (j-\sfrac12,j-\sfrac12) \ \ .  
\end{eqnarray}
The decomposition of atypical representations into modules of $\g^{(0)}$ 
can be found in eq.\ \eqref{evendec1}. In the special case of $j=0$ one 
obtains
$$ 
 [n] \otimes \P_\g(0) \  = \  \pi_\g \bigl([n]\bigr|_{\g^{(0)}} \otimes H_{(0)}
  \bigr) \ \oplus\  4 \cdot \P_\g(n) 
$$  
Here we have to insert the representation $H_{(0)} \ = \ 2(0,0) \oplus 2 
(\sfrac12,\sfrac12)$ of the bosonic subalgebra in the argument of $\pi_\g$.
\end{proposition}
\bigskip

From proposition 3 we may compute the tensor product of a projective 
cover with any other indecomposable through the following extension 
of formula \eqref{typInd} to projective covers of atypical 
representations, 
\begin{equation} 
\R \otimes \P_\g(j) \ \cong \ \S_\g(\R) \otimes \P_\g(j)  
\end{equation}
for all indecomposables $\R$. The symbol $\S$ has been introduced in 
equation \eqref{S} above such that it associates to $\R$ the direct 
sum of irreducibles that appear in its decomposition series. 
\medskip 

Note that so far we were able to express all tensor products through 
a direct sum of typical representations and projective covers. This 
is consistent with the before-mentioned general result that projective
representations form an ideal in the fusion ring of representations. 
There remains, however, one more family of tensor products to be 
determined. These are the tensor products between two atypical 
representations. Not surprisingly, we shall encounter a new set 
of indecomposables in such tensor products. These possess the 
following form 
\begin{eqnarray} \label{pi1}
\pi_{i \ti j}^{\rm indec} & = & 
  \bigoplus_{k = |i-j|}^{i+j-1} \ [k+\sfrac12] 
\longrightarrow  
  \bigoplus_{k = |i-j|}^{i+j} \ 2 [k]
\longrightarrow 
 \bigoplus_{k = |i-j|}^{i+j-1} \ [k+\sfrac12]  \ \ (i \neq j) 
\\[2mm] \label{pi2}   
 \pi_{j \ti j}^{\rm indec} & = & 
  \bigoplus_{k = 0}^{2j-1} \ [k+\sfrac12] 
\longrightarrow  3 [0] \ \oplus \ 
  \bigoplus_{k = 1}^{2j} \ 2[k]
\longrightarrow 
 \bigoplus_{k = 0}^{2j-1} \ [k+\sfrac12]  \ \ .
\end{eqnarray}  
Note that the second line is essentially a special case of the first 
except that we replaced the term $2[j-j]$ by $3[0]$. When rewritten in 
terms of our 2-dimensional diagrams, these representations read,  
{\xymatrixrowsep{4pt}
\xymatrixcolsep{4pt}
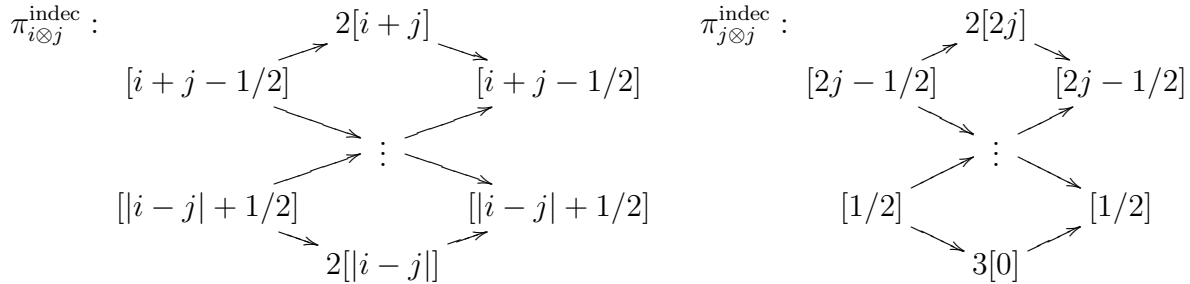
\begin{equation}\nonumber 
 \pi_{i \ti j}^{\rm indec}: \xymatrix{ &  2[i+j] \ar[dr]& \\
 [i+j-1/2]\ar[dr] \ar[ur] & & [i+j-1/2]\\
     & \ \vdots\ \ar[ur]\ar[dr] &  \\
              [|i-j|+1/2] \ar[ur]\ar[dr]&&[|i-j|+1/2] \\
 & 2[|i-j|] \ar[ur] &}\ \ \ \  
 \pi_{j \ti j}^{\rm indec}:
  \xymatrix{ &  2[2j] \ar[dr]& \\
 [2j-1/2]\ar[dr] \ar[ur] & & [2j-1/2]\\
     & \ \vdots\  \ar[ur]\ar[dr]&  \\
              [1/2] \ar[ur]\ar[dr]&&[1/2] \\
 & 3[0] \ar[ur] &}
\end{equation}}

\noindent 
{\bf Proposition 4:} {\it {\em (Tensor product of atypical representations)}  
The tensor product of two atypical representations $[i]$ and $[j]$ 
is given by 
\begin{equation} [i] \otimes [j] \ = \ \pi_\g\bigl((i,i)
  \otimes(j,j)\bigr)
   \ \oplus \ \delta_{i,j}\,  [0]\,  \oplus \ 
    \pi^{\rm indec}_{i\ti j} \ \ .
\end{equation} 
The formula holds for $i,j = 1/2,1,3/2,\dots$. Tensor products 
of the trivial representation $[0]$ with any other atypical 
representation are obvious.} 
\bigskip 
 
Let us note that for $i = j$ one copy of the atypical representation 
$[0]$ appears as a summand, i.e.\ it is not part of the single 
indecomposable representation that contains all the other atypical 
building blocks. We have also calculated a few tensor products 
between the new indecomposable representations $\pi_{i\ti j}$
and atypicals. Such products turn out to generate further 
indecomposables with a structure that is similar to the one of 
$\pi_{i\ti j}$ but involves different multiplicities. It seems
within reach to fully analyze the fusion ring that is generated by 
irreducibles, but since the applications we have in mind do not 
require such an exhaustive study, we have not investigated these 
issues much further.     
\bigskip

\subsection{Proof of the decomposition formulas - general ideas} 

There are two main ideas that enter into the proof of the above 
formulas. To begin with, we shall exploit systematically that the 
outer automorphisms are implementable not only in the irreducibles
of $\g$  but also in all their tensor products. This will ultimately 
organize all the typical subrepresentations and, more importantly, 
the elements of the composition series of indecomposables into 
multiplets of the group \SL2C of outer automorphisms. 

In a second step we then restrict the action of $\g$  to the action 
of an embedded $\h=\sll(2|1)$. Knowledge about the representation 
theory of the latter \cite{Marcu:1979se,Marcu:1979sg} (see also 
\cite{Frappat:1996pb}) along with a few new results from 
\cite{Gotz:2005jz} will then allow us to uniquely determine the 
structure of indecomposable $\g$   
representations that appear in the tensor products of irreducibles
and projectives. The consistency of our results has also been
verified by checking associativity of the tensor products involving
three representations of irreducible or projective type
with labels equal or below $4$ on a computer.%

\paragraph{Implementation of outer automorphisms.} We have stated 
above that the action of the \SL2C outer automorphisms of $\g$ can 
be implemented in all of its irreducible representations. Now we 
will argue that 
implementability of outer automorphisms is respected by the operation
of tensor products, by restriction to the socle and by quotients.
This implies that outer automorphisms may be implemented separately 
in all representations that appear in the decomposition of tensor 
products of irreducible representations. In this sense, such 
representations are rather special. 
\smallskip 

\begin{lemma}
Suppose that the action of outer automorphisms 
can be implemented in two representations $\pi$ and $\pi'$ of 
the Lie superalgebra $\g$ and that these implementations respect 
the $\Integer_2$ gradings of the representation spaces. Then it can also 
be implemented in the tensor product $\pi \otimes \pi'$.
\begin{proof}[\sc Proof:]
We denote the implementations of the outer automorphism 
$\c$ in the representations $\pi$ and $\pi'$ by $U_\pi(\c)$ and 
$U_{\pi'}(\c)$, respectively. Then the implementation in the tensor 
product is trivially given by $ U_{\pi\otimes \pi'}(\c) = U_\pi(\c) \otimes U_{\pi'}(\c)$. 
\end{proof} 
\end{lemma}

\begin{lemma}
Let $\pi$ be a representation of $\g$ in which 
the outer automorphism $\c$ is implemented by $U_\pi(\c)$. Let 
furthermore $\pi'$ be a subrepresentation that is invariant under 
the action of $U_\pi(\c)$, i.e.\ $U_\pi(\c){\cal H}_{\pi'}
\subset{\cal H}_{\pi'}$. Then the action of $\c$ is 
implementable in the factor representation $\pi/\pi'$.
\begin{proof}[\sc Proof:]
The statement is obvious. 
\end{proof}
\end{lemma}

\begin{lemma}
Suppose that the action of an outer automorphism
$\c$ can be implemented in a representation
$\pi$ of the Lie superalgebra 
$\g$ by the implementation map $U_\pi(\c):{\cal H}_\pi \rightarrow 
{\cal H}_\pi$ and let $\pi'$ be the socle of $\pi$. Then
$U_\pi(\c)\bigr|_{\mc{H}_{\pi'}}$ is an implementation of
$\gamma$ in $\pi'$, i.e.\ $U_\pi(\c)\mc{H}_{\pi'}\subset\mc{H}_{\pi'}$.
\begin{proof}[\sc Proof:] By definition, the socle ${\cal H}_{\pi'}$ 
is the maximal semisimple submodule of ${\cal H}_{\pi}$. Its image 
$U_\pi(\c) {\cal H}_{\pi'} \subset {\cal H}_\pi$ carries an action 
of the Lie superalgebra. The corresponding subrepresentation is 
given by $\pi' \circ \gamma$ and hence it decomposes into a sum 
of irreducibles just like $\pi'$ itself. In other words, the 
subspace $U_\pi(\c) {\cal H}_{\pi'}$ is a semisimple submodule 
of ${\cal H}_\pi$. From the maximality of the socle we therefore 
conclude ${\cal H}_{\pi'} \ = \ U_\pi(\c) {\cal H}_{\pi'}$.     
\end{proof} 
\end{lemma}

While the previous lemmas hold in full generality for all
Lie superalgebras, the next one relies on special properties of the
Lie superalgebra $\psl(2|2)$ and its class of \SL2C automorphisms
$\gamma_u$ as defined in \eqref{eq:Auto}.
\begin{lemma}
Let $\pi$ be a representation of $\g=\psl(2|2)$ in which an outer automorphism
$\c_u$ of the form \eqref{eq:Auto} is implemented by $U_\pi$ and let $\pi'$ be
an irreducible subrepresentation on the space $\mc{H}_{\pi'}$. Then $\pi$
defines a subrepresentation on the space $U_\pi\mc{H}_{\pi'}$ and the latter
is isomorphic to $\pi'$.
\begin{proof}
For any vector $v\in\mc{H}_{\pi'}$ we find
$\pi(x)U_\pi v=U_\pi\pi\circ\gamma^{-1}(x)v\subset U_\pi\mc{H}_{\pi'}$, hence
$U_\pi\mc{H}_{\pi'}$ is invariant. To see that the resulting
representation is isomorphic to the original one we just need to realize
a) that it is also irreducible, b) that the dimension agrees and c) that
the weights and weight multiplicities coincide. While b) is obvious, c) holds
because $\gamma_u$ acts trivially on the bosonic generators and especially
on the Cartan elements. Assume now
that there exists a proper subrepresentation on the space
$\mc{H}_{\pi''}\subset U_\pi\mc{H}_{\pi'}$.
Then, using the arguments of the first line above, we find a proper
subrepresentation of the original representation $\pi'$ on the space
$U_\pi^{-1}\mc{H}_{\pi''}\subset\mc{H}_{\pi'}$,
contradicting its irreducibility.
\end{proof}
\end{lemma}
\smallskip 

As simple as these statements are, they will be rather useful for 
our analysis of tensor products. They imply in particular, that 
all the indecomposable representations that arise in tensor 
products of irreducibles, and hence all the representations we
discuss in this note, allow for an implementation of the outer 
automorphisms. This insight is particularly useful when we analyze 
the internal structure of indecomposable composites. In fact, if 
the action of outer automorphisms is implementable in an 
indecomposable, then it is implementable in its socle and the 
associated factor representation. Hence, the transformation 
properties under the group \SL2C respect the structure of such 
indecomposable representations, i.e.\ they organize each floor 
of their decomposition diagram into \SL2C multiplets of
irreducible representations. For the indecomposables that 
appear in this text, the multiplicities are displayed 
in appendix \ref{ap:Multiplicities}.

\paragraph{Decomposition with respect to $\sll(2|1)$.} In this paragraph 
we shall study a particular embedding of $\h=\sll(2|1)$ into $\g$  
and explain how the irreducible representations of $\g$ decompose into 
representations of $\h$. We do not intend to present a complete 
introduction into $\sll(2|1)$ here, but restrict to a short list of 
relevant notations. More details can be found in the standard 
literature (see e.g.\ \cite{Frappat:1996pb}) and in 
\cite{Gotz:2005jz}. 
\smallskip 

The even subalgebra $\h^{(0)}$ of $\sll(2|1)$ is given by the sum 
$\gl(1) \oplus \sll(2)$. Consequently, the Kac modules are labeled by 
pairs $(b,j)$ where $b \in \Complex$ and $j = 1/2,1, \dots$. We 
shall denote these representations by 
$$  \p{b}{j} \ := \ 
    {\rm Ind}_{\h^{(0)} \oplus\, \h^{(1)}_-}^{\h} 
      V_{(b-1/2,j-1/2)} 
      \ = \ 
    \mc{U}(\h) \otimes_{\h^{(0)} \oplus\, \h^{(1)}_-} V_{(b-1/2,j-1/2)} \ \ .    
$$ 
The shift of the labels in our notations will turn out to be rather 
convenient in the following. We denote the Kac modules of $\h$ by 
brackets $\{\cdot,\cdot\}$ to distinguish them from those of $\g$. For 
$b\neq\pm j$, the Kac modules $\{b,j\}$ give rise to irreducible 
representations of dimension $d_{\{b,j\}} = 8j$. Kac modules 
$\{b,j\}$ with $b = \pm j$ are indecomposable. Their structure 
can be encoded in the following short diagram
$$ \{\pm j, j\}: \ 
   \{j\}_\pm \ \longrightarrow\  \{j-\sfrac12\}_\pm \ \ .  
$$ 
Here, $\{j\}_\pm$ denote the $(4j+1)-$dimensional 
irreducible atypical representations of $\h$. 
\smallskip 

The last type of representations that we shall need below are the 
projective covers ${\cal P}^\pm_\h({j})$ of the atypical representations 
$\{j\}_\pm$. For $j \neq 0$, their structure is encoded in the 
following picture 
\begin{equation} \label{sl21P}  
 {\cal P}^\pm_\h({j}):\ \{j\}_\pm \ \longrightarrow \ 
                       \{j+\sfrac12\}_\pm 
   \oplus \{j-\sfrac12\}_\pm \ \longrightarrow\ \{j\}_\pm \ \ . 
\end{equation} 
These spaces are $16j+4$ dimensional as one can easily check by adding 
up the dimensions of the atypical composition series. The projective 
cover of the trivial representation $\{0\}= \{0\}_\pm$ is an 
$8$-dimensional representation that is given by 
\begin{equation} \label{sl21P0} 
  {\cal P}_\h({0}): \ \{0\} \ \longrightarrow 
    \ \{\sfrac12\}_+ \oplus \{\sfrac12\}_- 
  \ \longrightarrow\ \{0\} \ \ .   
\end{equation} 
Results for tensor products of these representations are spelled 
out in appendix \ref{ap:sl21}. Tensor products of irreducible representations
were originally computed in \cite{Marcu:1979sg} and 
results for tensor products of irreducible with the projective 
representations ${\cal P}^\pm_\h(n)$ can be found in 
\cite{Gotz:2005jz}.%  
\smallskip 

What we shall need in the following are explicit formulas for the 
decomposition of irreducible $\g=\psl(2|2)$ representations with 
respect to the subalgebra $\h=\sll(2|1)$. Obviously, these 
depend on the explicit choice of the embedding. Here we shall 
consider the case in which the central element $Z \in \h^{(0)}$ 
of the even part in $\sll(2|1)$ is identified with the Cartan element 
of the first $\sll(2)$ subalgebra of $\g^{(0)}$. With this choice we find 
\begin{eqnarray} \label{atdec} 
[n]\bigr|_\h & \cong & \{n\}_+ \oplus  \{n\}_- \oplus \ 
     \mc{E}\bigl([n]\bigr) \\[2mm]  \nonumber 
  \text{where} & &  \mc{E}\bigl([n]\bigr) \ = \ 
 \bigoplus_{b=-n+1/2}^{n-1/2}\{ b , n + \sfrac12\} \ \ . 
\end{eqnarray}  
Below we shall think of $\mc{E}$ as a map that sends any sum 
of atypical $\g$-representations to a sum of typical representations
for $\h$. For typical representations $[i,j]$ of $\g$ we shall 
first assume that $j > i$. The other case will be treated below. 
\begin{equation}  \label{tdec} 
[i,j]\bigr|_\h \ \cong \ \bigoplus_{b=-i}^i \bigl( \{b,j+1\} \oplus 
                (1-\delta_{j,0}) \{b,j\} \oplus \{b-\sfrac12,j+\sfrac12\}
   \oplus \{b+\sfrac12,j+\sfrac12\}\bigr)\ \ .   
\end{equation}                  
Note that with our assumption $i<j$ only typical representations
occur on the right hand side. The formulas also holds true for 
$i>j$ and $i+j$ a half-integer. When $i>j$ and $i+j$ integer, 
on the other hand, the formula must be modified and the 
decomposition turns out to contain some of the projective 
covers ${\cal P}^\pm_\h(j)$ of atypical representations,  
\begin{eqnarray} 
[i,j]\bigr|_\h & \cong & {\bigoplus_{b=-i}^i}{}' \bigl( \{b,j+1\} \oplus 
           (1-\delta_{j,0}) \{b,j\} \oplus \{b-{\scriptstyle 
        \sfrac12},j+\sfrac12\}
   \oplus \{b+\sfrac12,j+\sfrac12\}\bigr) \ \oplus \nonumber
    \\[2mm] 
   & & \hspace*{3cm} 
 \ \oplus \, \bigoplus_{\nu = \pm} \bigl({\cal P}^\nu_\h({j}) \oplus 
         {\cal P}^\nu_\h({j+1/2})\bigr) 
\end{eqnarray} 
where the symbol $\oplus'$ instructs us to omit all terms that would 
formally be associated with atypical representations. In addition, 
we shall agree throughout this paper to replace the sum
${\cal P}^+_\h(0) \op {\cal P}^-_\h(0)$ by  ${\cal P}_\h(0)$
whenever it appears. This is 
purely formal and has no meaning in terms of representations. Note that 
the left hand side only contains projective representations. Let us 
finally spell out the decomposition formulas for the projective covers
and the indecomposables $\pi^{\text{indec}}$ which we defined in eq.\ (\ref{pi1},
\ref{pi2}), 
\begin{eqnarray} \label{Pdec} 
\P_\g(j)\bigr|_\h & \cong &   \bigoplus_{\nu = \pm} \bigl(\P^\nu_\h(j+\sfrac12) \, 
   \oplus\,  2 \cdot \P^\nu_\h(j)\, 
           \oplus\, \P^\nu_\h(|j-\sfrac12|)\bigr)\,  \oplus \ 
        \mc{E}\circ \S\bigl(\P_\g(j)\bigr)  \\[2mm]   
\pi^{\rm indec}_{i\ti j}\bigr|_\h & \cong & 
\bigoplus_{\nu = \pm} \, \bigl( \{|i-j|\}_\nu \, \oplus\, \{i+j\}_\nu
  \, \oplus \bigoplus_{p=|i-j|+\sfrac12}^{i+j+\sfrac12} 
\P_\h(p)\bigr) \ \oplus \  
\mc{E}\circ \S(\pi^{\rm indec}_{i\ti j})  \label{pidec}
\end{eqnarray} 
These formulas also hold for the special cases of $j=0$ and $i=j$
if we agree to replace $\{0\}_+ \oplus \{0\}_-$ by $\{0\}$. Once
more such a replacement is purely formal and has no meaning in terms
of representations. We note that it is rather easy to infer the 
projective covers $\P_\h$ in the two decomposition formulas from 
our planar pictures for $\P_\h$ and the indecomposables $\pi^{\text{indec}}$. 
This concludes the presentation of the background material that 
is needed in the proof of our decomposition formulas for 
tensor products of $\g$.

\subsection{Proof of the decomposition formulas - examples}

Rather than trying to go through the general proof of our formulas
we would like to illustrate the main ideas in two rather representative 
examples. These are the tensor product of the atypical representation
$[1]$ with itself and with $[0,2]$.  

\paragraph{The tensor product $[1]\otimes [1]$.}
Let us begin by collecting a few results on the 34-dimensional atypical 
representation $[1]$ of $\g$. With respect to the embedded $\h$, this 
representations decomposes as follows 
$$ [1]\bigr|_{\h} \ \cong \ \{1\}_+ \op \{1\}_- \op 
        \p{\sfrac12}{{\scriptstyle \frac{3}{2}}} \op  
        \p{-\sfrac12}{{\scriptstyle \frac{3}{2}}} \ \ . 
$$ 
Using standard results (see appendix \ref{ap:sl21}) about tensor products of 
irreducible representations of $\h$, we obtain the following 
decomposition formula for the tensor product $[1] \otimes [1]$ 
in terms of representations of $\h$, 
\begin{equation} \label{form1}  
 \bigl([1] \otimes [1]\bigr)\bigr|_{\h} \ \sim \ 2\cdot \{0\} \oplus 
  \ 2 \cdot {\cal P}_\h(0) \op \ 
      \bigoplus_{\nu=\pm} \ \bigl(\{2\}_\nu \oplus 
     3 \cdot {\cal P}^\nu_\h(\sfrac12) \oplus {\cal P}^\nu_\h(1) \oplus
     2 \cdot {\cal P}^\nu_\h({\scriptstyle \frac{3}{2}})\bigr) 
   \ + \ \dots 
\end{equation}  
where the dots stand for a sum of typical $\h$ representations. 
The latter will not play any role for the following analysis. 
\smallskip 

It is also easy to find the typical $\g$ representations in the 
tensor product of $[1]$ with itself, 
\begin{eqnarray} 
\bigl([1] \otimes [1]\bigr)^{\rm typ} & = & 
     [0,1]\op [0,2]\op [1,0] \op [1,2] \op [2,0] \op [2,1] 
  \nonumber \\[3mm]  
\bigl([1] \otimes [1]\bigr)^{\rm typ}\bigr|_{\h} & \sim &  
    \ 2 \cdot {\cal P}_\h(0) \op \ 
      \bigoplus_{\nu=\pm} \ \bigl(
     2 \cdot {\cal P}^\nu_\h(\sfrac12) \oplus {\cal P}^\nu_\h(1) \oplus 
      {\cal P}^\nu_\h({\scriptstyle \frac{3}{2}})\bigr) 
   \ + \ \dots \label{form2} 
\end{eqnarray} 
When we passed to the second line, we have inserted the results 
from our decomposition formulas for typical $\g$ representations 
with respect to the embedded $\h$. Once more, we have omitted 
all typical $\h$ representations. 
\smallskip 

A comparison of eqs.\ \eqref{form1} and \eqref{form2} gives the 
following intermediate result,
$$  \Bigl([1] \otimes [1] \ - \ 
     \bigl([1] \otimes [1]\bigr)^{\rm typ}\Bigr)\Bigr|_{\h}
    \ \sim \   2 \cdot \{0\} \oplus \bigoplus_{\nu=\pm} \ \bigl(
      \{2\}_\nu \oplus {\cal P}^\nu_\h(\sfrac12) \oplus  
      {\cal P}^\nu_\h({\scriptstyle \frac{3}{2}})\bigr)\ + \ \dots 
$$
The atypical $\h$ representations that remain must come 
from the decomposition of the $\g$ indecomposables that appear 
in the tensor product of $[1]$. A short glance on our decomposition 
formulas \eqref{atdec} and \eqref{pidec} confirms that this is 
indeed the case.  
\smallskip 

We can actually use this example to derive the structure of the
indecomposable representation $\pi_{1\ti 1}$ from the information 
on its restriction to $\h$. In fact, it is rather easy to see that  
the composition series of $\pi_{1\ti 1}$ contains the following 
list of atypical representations, each displayed with a multiplicity 
that refers to its transformation properties under the action of the 
\SL2C outer automorphisms\footnote{The subscripts may be determined 
from the \SL2C transformation properties of the bosonic multiplets
in the involved tensor factors. The latter are listed in appendix 
\ref{ap:Multiplicities}.}   
$$ [0]_1 \ , \ [0]_3 \ , \ 2 [\sfrac12]_1 \ , \ [1]_2 
     \ , \  2 [{\scriptstyle \frac{3}{2}}]_1 \ , \ [2]_2\ \ . 
$$ 
A moment of thought reveals that all but the $[0]_1$ representation
must be part of one single indecomposable in order to be able to recover 
our knowledge about the decomposition with respect to $\h$. It is at 
this point where our assignment of multiplicities becomes crucial. In 
fact, our knowledge from the $\h$ embedding would e.g.\ have been 
consistent with including two of the four trivial $\psl(2|2)$ representations 
into the indecomposable. But since there is no doublet, we were forced 
to include the triplet. Hence, there is only a single irreducible 
representation $[0]$ left. In other words, the tensor product 
$[1] \otimes [1]$ contains only one true invariant. The presence of
additional states which transform trivially but sit in the indecomposables
has to be contrasted with the case of ordinary simple Lie algebras where the
tensor product of a representation with its conjugate contains precisely one
such state.

\paragraph{The tensor product $[1] \otimes [0,2]$.}
With respect to the embedded $\h \subset \g$, the typical $\g$ representation 
$[0,2]$ decomposes as follows 
$$ [0,2]\bigr|_{\h}  \ \cong \ \p{0}{3} \op \p{0}{2} \op 
        \p{\sfrac12}{{\scriptstyle \frac{5}{2}}} \op  
        \p{-\sfrac12}{{\scriptstyle \frac{5}{2}}} \ \ . 
$$ 
Using again standard results about tensor products of irreducible 
representations of $\h$ we obtain the following decomposition 
formula for the tensor product $[1] \otimes [0,2]$ in terms of
representations of $\h$, 
\begin{equation} \label{form3}  
 \bigl([1] \otimes [0,2]\bigr)\bigr|_{\h} \ \sim \ 
      \bigoplus_{\nu=\pm} \ \bigl(
     {\cal P}^\nu_\h(\sfrac12) \oplus 2 \cdot {\cal P}^\nu_\h(1) \oplus
     {\cal P}^\nu_\h({\scriptstyle \frac{3}{2}})\bigr) 
   \ + \ \dots 
\end{equation}  
where the dots stand for a sum of typical $\h$ representations
as in the last example. One can easily find the typical $\g$
representations in the tensor product of $[1]$ with $[0,2]$, 
\begin{eqnarray} 
\bigl([1] \otimes [0,2]\bigr)^{\rm typ} & = & 
     [{\scriptstyle \frac{1}{2}},{\scriptstyle \frac{3}{2}}]\op  
[{\scriptstyle \frac{1}{2}},{\scriptstyle \frac{5}{2}}]\op
 [{\scriptstyle \frac{1}{2}},{\scriptstyle \frac{7}{2}}]
\op 2[1,2] \op 2[1,3] \op [{\scriptstyle \frac{3}{2}},
{\scriptstyle \frac{5}{2}}] \ \ . \label{form4} 
\end{eqnarray} 
We observe that none of these representations contributes any 
indecomposable upon restriction to the embedded $\h$. Hence, 
we obtain 
\begin{equation}\label{102dec}  
  \Bigl([1] \otimes [0,2] \ - \ 
     \bigl([1] \otimes [0,2]\bigr)^{\rm typ}\Bigr)\Bigr|_{\h} 
    \ \sim \   \bigoplus_{\nu=\pm} \ \bigl(
      {\cal P}^\nu_\h({\sfrac12}) \oplus  
      2 \cdot {\cal P}^\nu_\h({1}) \oplus
       {\cal P}^\nu_\h({3/2})  \bigr)\
+
\ \dots 
\end{equation} 
According to eq.\ \eqref{Pdec}, this particular sum of  
projective $\h$ representations comes from the decomposition 
of the projective cover $\P_\g(1)$. Hence, we conclude that 
the latter appears in the tensor product of $[1]$ with $[0,2]$, 
in agreement with our proposition 1.  
\smallskip 

Before we conclude, we would like to point out that the we 
can read off the internal structure of $\P_\g(1)$ from the 
information on its restriction to $\h$. In fact one can see
rather easily that the composition series of the indecomposables 
in the tensor product of $[1]$ and $[0,2]$ contains the 
following list of atypical representations, each displayed 
with a multiplicity that refers to its transformation properties
on the action of the outer automorphisms,\footnote{The relevant 
data that allow to determine the mutiplicities can be found in 
appendix \ref{ap:Multiplicities}.}
$$ [0]_2  \ , \ 2 [\sfrac12]_2 \ , \ [1]_3 \ , \
   3 [1]_3 \ , \  2 [{\scriptstyle \frac{3}{2}}]_2 \ , 
\ [2]_1\ \ . 
$$ 
Our planar picture for $\P_\g(1)$ provides the unique pattern 
in which we can form a composite from these constituents that 
is consistent with the information \eqref{102dec} on the 
restriction to the subalgebra $\h$. 

\section{Conclusions and Outlook} 

In this note we have succeeded to decompose all tensor products 
between finite dimensional irreducible and projective representations
of $\psl(2|2)$. Whereas tensor products involving at least one 
projective representation were shown explicitly to stay within the 
class of projectives, we have constructed a new family of finite 
dimensional indecomposable representations that appear in the 
tensor product of atypicals. Preliminary investigations show that 
tensor products of these new indecomposables $\pi_{i\ti j}$ with 
atypicals generate yet another family of representations whose 
structure resembles the one of $\pi_{i\ti j}$, though with 
different multiplicities of the involved atypical building 
blocks. Since the applications we have in mind only require  
tensor products in which at least one factor is projective, 
we have not pushed our investigations further into this 
direction. We believe, however, that results can be obtained 
using the techniques we have developed above. Even though 
indecomposables of $\psl(2|2)$ cannot be classified, it may 
well be possible to classify all those representations that 
arise in multiple tensor products of irreducibles. In fact, 
according to section 3.2, the latter admit an implementation 
of the \SL2C outer automorphisms and therefore they form a rather 
distinguished sub-class of representations. 
\smallskip 

The techniques we have used here may also be applied to other 
Lie superalgebras of the A-series, in particular to $\psl(n|n)$. 
A promising approach would be to address $\sll(n|1)$ first and 
then to advance to non-trivial second label. Obviously, the 
structure of the representation theory, becomes much richer for
larger Lie superalgebras, in particular because multiply atypical
representations can occur (see, e.g., \cite{Hughes2000:MR1765833}
and references therein). Some partial results in this direction 
will be published elsewhere. 
\medskip 

We finally want to sketch at least one concrete physics problem 
to which we hope to apply the rather mathematical results of this
note. During recent years, non-linear \mbox{$\sigma$-models} on supergroups 
and supercosets have surfaced in a variety of distinct problems and 
in particular through studies of string theory in certain RR 
backgrounds \cite{Metsaev:1998it,Rahmfeld:1998zn,Berkovits:1999im}. 
Many specific and important properties of these models, such as 
e.g.\ the possible existence of conformal invariance even in the absence 
of a Wess-Zumino term, originate from peculiar features of the 
underlying Lie superalgebra \cite{Bershadsky:1999hk,
Berkovits:1999zq}. 
\smallskip 

Since the isometries of $AdS_3\times S^3$ are generated by two copies
of the even subalgebra $\sll(2,\Real)\oplus\mf{su}(2)$
of the non-compact real form $\mf{psu}(1,1|2)$ of $\psl(2|2)$,
it is not hard to believe that the corresponding 
$\sigma$-models enter the description of strings in an $AdS_3$ 
background \cite{Rahmfeld:1998zn,Berkovits:1999im}. In fact,
for strings moving in the presence of a pure NSNS background field
the physics is described by a WZW model for $\psl(2|2)$. This 
theory possesses a holomorphic and antiholomorphic
$\psl(2|2)$ current symmetry and it may be 
solved exactly after decoupling \cite{Bars:1990hx,Berkovits:1999im}
the bosons and the fermions, using results established in
\cite{Teschner:1997ft,Maldacena:2000hw,Kausch:2000fu,Ponsot:2001gt,
Maldacena:2001km} and references therein. After the marginal deformation
which arises from turning on a RR background field, however,
the local (worldsheet) symmetries 
of the system are reduced drastically \cite{Bershadsky:1999hk} 
and so far no solution has been found, in spite of the 
significant interest in such models (see also e.g.\ 
\cite{Zirnbauer:1999ua}). 
\smallskip 

In a first step one may hope to determine the exact spectra of 
theories with $AdS_3\times S^3$ target as a function of the
strength of the RR 
flux. Results in \cite{Bershadsky:1999hk} imply that states 
which transform according to the same representation of the 
remaining global $\psl(2|2)\oplus\psl(2|2)$ symmetry experience the same energy shift as we 
switch on the RR background field. This motivates to 
classify all string states according to their behavior under 
the action of $\psl(2|2)$. Such states arise by application of 
creation operators on certain ``ground states'' in the theory. At 
the WZW-point, these creation operators are the negative modes of the 
$\psl(2|2)$ currents. The latter transform in the adjoint 
representation of $\psl(2|2)$. Hence, in order to determine 
the transformation properties of excited states, we must 
control tensor powers (and the symmetric parts therein) of 
the adjoint representation. This is exactly where the results 
of the present note feed into studies of strings in $AdS_3$.  
Such an analysis is beyond the scope of this work but we plan 
to come back to these issues in a forthcoming publication.         
\bigskip 
\bigskip

\noindent 
{\bf Acknowledgment:} It is a pleasure to thank Gleb Arutyunov, 
Jerome Germoni, Hubert Saleur, Paul Sorba and Anne Taormina for 
many useful discussions. This 
work was partially supported by the EU Research Training Network grants 
``Euclid'', contract number HPRN-CT-2002-00325, ``Superstring Theory", 
contract number MRTN-CT-2004-512194, and ``ForcesUniverse'', contract 
number MRTN-CT-2004-005104. TQ is supported by a PPARC postdoctoral
fellowship under reference PPA/P/S/2002/00370 and partially by the
PPARC rolling grant PPA/G/O/2002/00475. We are grateful for the kind
hospitality at the ESI during the workshop ``String theory in curved
backgrounds'' which stimulated the present work.%
\newpage

\appendix
\section{\label{ap:Multiplicities}Appendix: \SL2C multiplicities}
\setcounter{equation}{0}

As it is mentioned in the main text, the implementation of the outer 
automorphisms helps a lot in understanding the decomposition of tensor 
products. In this appendix we would like to explain how the \SL2C
action organizes representations into multiplets and list explicit 
results for all the indecomposables we are interested in. 
\smallskip 

Let us recall that the \SL2C automorphisms act trivially on the 
bosonic subalgebra. Hence, each representation in which these 
automorphisms are implemented may be decomposed into a sum of 
$$ (i,j) \otimes V^J \ \cong \ (i,j)_n     \ \ \ \ \text{ where } 
   \ \ \ n \ = \ 2J+1\ \ , $$       
$(i,j)$ are representations of the even subalgebra and $V^J$ carries
an action of the automorphism group \SL2C. Since all our representations
are assumed to be finite dimensional, the same must be true for $V^J$.
This means that only \SL2C representations with half-integer spin $J$ 
and dimension $n=2J+1$ can arise.
\smallskip 

With the previous remarks in mind we can now move ahead and 
analyse how the various representations that appeared in the 
main text decompose into the building blocks $(i,j)_n$. We 
shall restrict our explicit lists here to the irreducible 
representations. Let us begin with the generic typical 
irreducible representations for $i,j>1/2$:
\eqn\label{Multi1}
[i,j] \ = \ \begin{array}{lll}
&                                     &\quad(i+1,j)_1 \\
                    &\quad (i+\sfrac12,j+\sfrac12)_2 & \\
                   &                   &\quad(i,j+1)_1\\
                    &\quad(i+\sfrac12,j-\sfrac12)_2   &\\
  (i,j)_3                  &   &   \qquad     (i,j)_1 \\
                  &  \quad(i-\sfrac12,j+\sfrac12)_2&\\
                                    && \quad (i-1,j)_1\\
                  & \quad (i-\sfrac12,j-\sfrac12)_2  &\\
                                     &&\quad (i,j-1)_1\ \ .
\end{array}
\qen
In case one of the labels $i,j$ is equal to $1/2$, the 
generic decomposition gets reduced. When $j=1/2$ and 
$i>1/2$ we find 
\eqn
[i,\sfrac12] \ = \ \begin{array}{lll} 
                    &\quad(i+\sfrac12,1)_2 &\quad(i+1,\sfrac12)_1 \\
                   &                   &\\
                    &\quad(i+\sfrac12,0)_2   &\qquad(i,\tfrac32)_1\\
  (i,\sfrac12)_3                  &   &     \\
                  & \quad (i-\sfrac12,1)_2& \qquad   (i,\sfrac12)_1\\
                                    &&\\
                  &\quad  (i-\sfrac12,0)_2  &\quad (i-1,\sfrac12)_1\ \ .
\end{array}
\qen
Obviously, the case of $i=1/2$ and $j>1/2$ is analogous. The series 
of representations with $j=0, i>1/2,$ possesses an even shorter picture 
\eqn
[i,0] \ =\ \begin{array}{lll}
&                                     &\quad (i+1,0)_1 \\
                    &\quad(i+\sfrac12,\sfrac12)_2 & \\
  (i,0)_3                  &   &  \qquad(i,1)_1        \\
                  & \quad (i-\sfrac12,\sfrac12)_2&\\
                                     &&\quad (i-1,0)_1 \ \ . 
\end{array}
\qen
The last typical representation that remains to be 
treated is the case of $i=1/2$ and $j=0$ for which one finds 
\eqn
[\sfrac12,0]\ =\ \begin{array}{lll}
                     &\quad(1,\sfrac12)_2 &\quad (\tfrac32,0)_1 \\
  (\sfrac12,0)_3                  &   &         \\
                  & \quad (0,\sfrac12)_2&\quad (\sfrac12,1)_1\ \ .
\end{array}
\qen
Now we can turn to the irreducible atypical 
representations with $j \geq 1/2$, 
\eqn
[j] \ =\ \begin{array}{ll}
 & \quad (j+\sfrac12,j-\sfrac12)_1 \\
                (j,j)_2  &\\
 & \quad (j-\sfrac12,j+\sfrac12)_1\ \ .
\end{array}
\qen
The representation $[0]$ is trivially given by $(0,0)_1$. This 
concludes our list of irreducible representations. Similarly, 
we could now analyse all those indecomposables which allow for
an implementation of the \SL2C automorphisms. Let us stress, 
that the requirement of implementability is not fulfilled for 
the Kac modules $[j,j]$. Hence, the above formulas should only 
be used for $i \neq j$. 
\medskip 

In subsection 3.2 we argued that the atypical constituents of 
the indecomposables $\P_\g(j)$ and $\pi_{i\ti j}$ are organized 
in multiplets of \SL2C. For such multiplets we shall employ the 
symbol $[j]_m$. Note that the decomposition of $[j]_m$ in terms
of $\g^{(0)} \oplus \sl2$ representations is obtained from the 
corresponding decomposition of $[j]$ (see above) by tensoring 
with the \SL2C representation $V^I$ where $m = 2I+1$.  
In case of the projective covers $\mathcal{P}
(j), j \neq 0,$ one finds the following structure,
 \begin{equation} \P_\g(j): \ 
  \xymatrix{& &  [j+1]_1 \ar[dr]& &\\
 & [j+1/2]_2\ar[dr] \ar[ur] & & [j+1/2]_2\ar[dr]&\\
     [j]_1\ar[dr] \ar[ur] && [j]_{3,1}\ar[ur]\ar[dr] & &[j]_1\ . \\
             & [j-1/2]_2 \ar[ur]\ar[dr]&&[j-1/2]_2\ar[ur] &\\
& & [j-1]_1 \ar[ur] &}
\end{equation}
By $[j]_{3,1}$ we mean that there is one triplet $[j]_3 $ and 
one singlet $[j]_1$. This result can be verified by the explicit 
decomposition of the tensorproduct between $[1/2]$ and $[j+1/2,
j-1/2]$ in which, up to typicals, exactly one projective cover 
$\mathcal{P}(j)$ appears.  
\smallskip 

$\mathcal{P}(0)$ has to be treated seperately. Its structure is 
encoded in a picture of the form 
 \begin{equation} \P_\g(0): \ 
  \xymatrix{& &  [1]_2 \ar[dr]& &\\
 & [1/2]_3\ar[dr] \ar[ur] & & [1/2]_3\ar[dr]&\\
     [0]_1 \ar[ur] && [0]_{5,1}\ar[ur] & &[0]_1\ . }
 \end{equation}
Finally, we also want to list the multiplicities for the 
indecomposables $\pi_{i \ti j}$ that arise in tensor products
of atypicals. For these representations we find  
{\xymatrixrowsep{4pt}
\xymatrixcolsep{4pt}
\begin{equation}\nonumber 
 \pi_{i \ti j}^{\rm indec}: \xymatrix{ &  [i+j]_2 \ar[dr]& \\
 [i+j-1/2]_1\ar[dr] \ar[ur] & & [i+j-1/2]_1\\
     & \ \vdots\ \ar[ur]\ar[dr] &  \\
              [|i-j|+1/2]_1 \ar[ur]\ar[dr]&&[|i-j|+1/2]_1 \\
 & [|i-j|]_2 \ar[ur] &}\ \ 
 \pi_{j \ti j}^{\rm indec}:
  \xymatrix{ &  [2j]_2 \ar[dr]& \\
 [2j-1/2]_1\ar[dr] \ar[ur] & & [2j-1/2]_1\\
     & \ \vdots\  \ar[ur]\ar[dr]&  \\
              [1/2]_1 \ar[ur]\ar[dr]&&[1/2]_1 \\
 & [0]_3 \ar[ur] &}
\end{equation}}

\section{\label{ap:sl21}Appendix: Tensor products for $\sll(2|1)$}

In order to list results on the tensor products of $\sll(2|1)$ 
representations, we would like to introduce a map $\pi_\h$ 
which sends representations of the bosonic subalgebra $\h^{(0)}$ 
to typical representations of $\h$. Its action on irreducibles
is given by 
\begin{equation}
\pi_\h(b-\sfrac12,j-\sfrac12) \ = \  
 \begin{cases} \{b,j\} & \text{ for } \ \ b \neq \pm j \ \ , 
                   \\[2mm] 
                   0 & \text{ for } \ \ b = \pm j \ \ .  
\end{cases} 
\end{equation} 
The map $\pi_\h$ may be extended to a linear map on the space of 
all finite dimensional representations of $\h^{(0)}$. 
\smallskip 

The first tensor product we would like to display is the one 
between two typical representations \cite{Marcu:1979sg}. In 
our new notations, the decomposition is given by 
\begin{eqnarray} 
 \{b_1,j_1\}\otimes\{b_2,j_2\} & = &  
   \pi_\h\bigl((b_1-\sfrac12,j_1-\sfrac12)\otimes {\{b_2,j_2\}\bigr|_{\h^{(0)}}}\bigr)
      \  \oplus\ \\[4mm] \nonumber & & \hspace*{-4cm} \oplus \  
    \left\{ \begin{array}{cl} 
   \mc{P}_\h (\pm |b_1+b_2|\mp \sfrac12) &  \text{for} \  
                  b_1+b_2\ =\ \pm(j_1+j_2) \\[2mm] 
\mc{P}_\h^\pm(|b_1+b_2|)\oplus\mc{P}_\h^\pm(|b_1+b_2|-\sfrac12)
  & \text{for} \ b_1+b_2\in \pm
  \{|j_1-j_2|+1,\cdots,j_1+j_2-1\} \\[2mm] 
      \mc{P}_\h(\pm |b_1+b_2|) & \text{for} \  
    b_1+b_2 \ = \ \pm |j_1-j_2|\ \ .  
\end{array} \right. 
\end{eqnarray}    
Note that neither $j_1$ nor $j_2$ can vanish so that the three 
cases listed above are mutually exclusive. The first term computes
all the typical representations that appear in the tensor product. 
All it requires is the decomposition of typical $\h$ representations 
into irreducibles of the bosonic subalgebra, 
$$ \{b,j\}\bigr|_{\h^{(0)}} \ = \ (b,j) \oplus (b+\sfrac12,j-\sfrac12) \oplus 
                (b-\sfrac12,j-\sfrac12) \oplus (b,j-1)\ \ . 
$$ 
and a computation of tensor products for representations of 
$\h^{(0)} = \gl(1) \oplus \sll(2)$ which presents no difficulty. 
The outcome is then converted into a direct sum of typical 
representations through our map $\pi_\h$.  
\smallskip 

Tensor products of typical with atypical representations can also
be found in Marcu's paper. The results are 
\begin{eqnarray} 
 \{b_1,j_1\}\otimes\{j_2\}_\pm & = &  
   \pi_\h\bigl( (b_1-\sfrac12,j_1-\sfrac12)\otimes \{j_2\}_\pm\bigr|_{\h^{(0)}} \bigr)
      \  \oplus\ \\[4mm] \nonumber & & \hspace*{-2cm} \oplus \  
    \begin{cases} 
   \mc{P}^\mp_\h (|b_1\pm j_2|- \sfrac12) & \ \ \ \ \text{for} \ \  
                  b_1\pm j_2\ \in \ \mp \{|j_1-j_2|+1, \dots, j_1+j_2 \} 
    \\[2mm] 
      \mc{P}^\pm_\h(|b_1\pm j_2|) & \ \ \ \ \text{for} \ \  
    b_1\pm j_2 \ \in \ \pm \{|j_1-j_2|, \dots, j_1+j_2-1\} \ .  
\end{cases} 
\end{eqnarray}    
This formula may be used to determine the tensor product of 
typical representations with projective covers and other 
indecomposables. These tensor products are simply given by 
\begin{equation} \label{TX} 
\{b,j\} \otimes \mc{R} \ \cong \ \{b,j\} \otimes 
          \S_\h(\mc{R}) \ \ \ \ .  
\end{equation} 
Here, the maps $\S_\h$ is defined such that it sends the 
indecomposable representation $\mc{R}$ to a direct sum of its
atypical building blocks, i.e.\ 
\begin{equation} \label{Sh}   
\S_\h(\mc{R})  \ = \ \bigoplus_{\pm j} \ [\mc{R}:\{j\}_\pm]\ 
 \{j\}_\pm\ \ .  
\end{equation} 
Here, $[\mc{R}:\{j\}_\pm]$ counts how many times the atypical 
representation $\{j\}_\pm$ appears in the composition series of
$\mc{R}$. In the special case of $\mc{R} = \P_\h^\pm(j)$, 
these numbers may be read off from the diagrams \eqref{sl21P}
and \eqref{sl21P0}. 
\smallskip 

Having gone through the entire list of products which involve
at least one typical factor, we would like to turn to the fusion 
of projective covers with any other type of representation. The 
tensor product between a projective cover $\P^\pm_\g(j)$
and an atypical representation $\{n\}_\pm$ with $n>0$ is given by  
\begin{eqnarray} \nonumber 
 \P^\pm_\h(j) \otimes \{n\}  & = &  
   \pi\bigl(H^\pm_j \otimes \{n\}\bigr|_{\h^{(0)}}\bigr)
\ \oplus \ \P_\h(\pm j + n) \ \\[2mm] 
\text{ where } & & H^\pm_j \ = \ (\pm j -\sfrac12,j-\sfrac12) \oplus 
(\pm (j +\sfrac12)-\sfrac12,j)  \label{Hpmj}
\end{eqnarray}    
and $H^\pm_0 = H_0 = (0,0)\oplus(-1,0)$. We also set $\{|n|\}_\pm=
\{\pm |n|\}$. 
As before, we can exploit this formula further to determine all 
tensor products between a projective cover and an indecomposable 
composite of atypical representations. The corresponding formula
employs our symbol $\mc{S}_\h$ (see eq.\ \eqref{Sh}), 
\begin{equation} \label{PX} 
\P_\h^\pm(j) \otimes \mc{R} \ \cong \ \P_\h^\pm(j) \otimes 
          \S_\h(\mc{R}) \ \ \ \ .  
\end{equation} 
As an application, we can spell out the tensor product between two 
projective covers $\P^\pm_\h(j_1), j_1 \geq 0,$ and $\P_\h(j_2) = 
\P^{\sign{j_2}}_\h(|j_2|)$, 
\begin{eqnarray} 
 \P^\pm_\h(j_1) \otimes \P_\h(j_2) & = &  
   \pi_\h\bigl(H^\pm_{j_1}  
        \otimes \P_\h(j_2)\bigr|_{\h^{(0)}}\bigr)
      \  \oplus\ \\[4mm] \nonumber & & 
\hspace*{-1cm} \oplus \  
\P_\h(\pm j_1+ j_2+\sfrac12) \oplus 2\cdot \P_\h(\pm j_1+ j_2) \oplus 
\P_\h(\pm j_1+ j_2-\sfrac12) \ \ . 
\end{eqnarray}    
The $\g^{(0)}$ modules $H^\pm_j$ were defined in eq.\ \eqref{Hpmj}. 
In the argument of $\pi_\h$ the product $\otimes$ refers to the 
fusion between representations of the bosonic subalgebra 
$\h^{(0)} = \gl(1) \oplus \sll(2)$.
\smallskip 

We are finally lacking a formula for the tensor product of two 
atypical representations. According to \cite{Marcu:1979sg}, such 
products are given by 
\begin{eqnarray} 
  \{j_1\}_\pm\otimes\{j_2\}_\pm
  & = &  \{j_1+j_2\}_\pm\ \oplus\ 
 \bigoplus_{j=|j_1-j_2|}^{j_1+j_2-1}\bigl\{\pm(j_1{+}j_2{+}\sfrac12),j+\sfrac12\bigr\}\ \ , 
\\[2mm] 
\{j_1\}_+\otimes\{j_2\}_-
  & = &  \bigl\{|j_1-j_2|\bigr\}_{\sign(j_1-j_2)}\ \oplus\ 
  \bigoplus_{j=|j_1-j_2|+1}^{j_1+j_2}\{j_1{-}j_2,j\}\ \ .
\end{eqnarray} 
Proofs for all these formulas can be found in \cite{Marcu:1979sg}
and in our recent paper \cite{Gotz:2005jz}. 

\def\cprime{$'$} \def\cprime{$'$}
\providecommand{\href}[2]{#2}\begingroup\raggedright\endgroup

\end{document}